# $\mathcal{CP}$-Violation and the Quark Mass Matrices

B. Margolis [1] and S. Punch

*Department of Physics, McGill University, 3600 University St.,*
*Montréal, Québec, Canada H3A 2T8*

C. Hamzaoui

*Département de Physique, Université du Québec à Montréal,*
*Case Postale 8888, Succ. Centre-Ville,*
*Montréal, Québec, Canada H3C 3P8*

## Abstract

We study a class of quark mass matrix models with the $\mathcal{CP}$-violating phase determined through making the $\mathcal{CP}$-violating parameter $J$ an extremum. These models assume that $m_u \ll m_c \ll m_t$ and that $m_d \ll m_s \ll m_b$. They have $\left|\frac{V_{ub}}{V_{cb}}\right| \approx \sqrt{\frac{m_u}{m_c}}$, $\left|\frac{V_{td}}{V_{ts}}\right| \approx \sqrt{\frac{m_d}{m_s}}$ and $|V_{us}| \approx |V_{cd}| \approx \sqrt{\frac{m_d}{m_s} + \frac{m_u}{m_c}}$. The Wolfenstein parameters $\rho$ and $\eta$ are found to be related by $\rho \approx \eta^2$. Finally, we examine a special class of such models where the masses are constrained to be roughly in geometric progression. Further application of the extremal condition to $J$ then leads to $\sqrt{\frac{m_d}{m_s}} \approx 3\sqrt{\frac{m_u}{m_c}}$ and hence $\eta \approx \frac{1}{3}$, for maximal $J$.

## The Models

We study here the models of the quark mass matrices of the type

$$M_U = \Lambda_U \begin{pmatrix} 0 & A & 0 \\ A^* & B & C \\ 0 & C & 1 \end{pmatrix} \quad , \quad M_D = \Lambda_D \begin{pmatrix} 0 & D & 0 \\ D^* & E & F \\ 0 & F & 1 \end{pmatrix} \quad (1)$$

which incorporate in a characteristic manner the hierarchy $m_u \ll m_c \ll m_t$ and $m_d \ll m_s \ll m_b$. We can take all matrix elements as real except either $A$ or $D$, and we choose to make $D$ complex. Then for $|D|^2 \ll |E - F^2| \ll 1$

$$\frac{m_d}{\Lambda_D} \approx \frac{|D|^2}{|E-F^2|} \qquad \frac{m_s}{\Lambda_D} \approx |E - F^2| \qquad \frac{m_b}{\Lambda_D} \approx 1 \quad (2a)$$

and for $A^2 \ll |B - C^2| \ll 1$

$$\frac{m_u}{\Lambda_U} \approx \frac{A^2}{|B-C^2|} \qquad \frac{m_c}{\Lambda_U} \approx |B - C^2| \qquad \frac{m_t}{\Lambda_U} \approx 1 \ . \quad (2b)$$

---

[1] Presented by B.M. at the MRST Meeting, Rochester, New York, May 8, 1995

We have then

$$\tilde{a}^2 \equiv \left(\frac{A}{B-C^2}\right)^2 \approx \frac{m_u}{m_c} \quad \text{and} \quad \tilde{b}^2 \equiv \left(\frac{|D|}{E-F^2}\right)^2 \approx \frac{m_d}{m_s} \tag{3}$$

and taking $A$ to be positive

$$A \approx \frac{\sqrt{m_u m_c}}{m_t} \quad \text{and} \quad |D| \approx \frac{\sqrt{m_d m_s}}{m_b} . \tag{4}$$

Our mass matrices then have the approximate form

$$M_U = \begin{pmatrix} 0 & \sqrt{m_u m_c} & 0 \\ \sqrt{m_u m_c} & B m_t & C m_t \\ 0 & C m_t & m_t \end{pmatrix}, M_D = \begin{pmatrix} 0 & \sqrt{m_d m_s}\, e^{i\beta} & 0 \\ \sqrt{m_d m_s}\, e^{-i\beta} & E m_b & F m_b \\ 0 & F m_b & m_b \end{pmatrix} \tag{5a}$$

with

$$|B - C^2| \approx \frac{m_c}{m_t} \quad \text{and} \quad |E - F^2| \approx \frac{m_s}{m_b} . \tag{5b}$$

The quark mixing matrix is given by the approximate expressions

$$V_{ud} \approx \frac{1}{\sqrt{1+\bar{a}^2}\sqrt{1+\bar{b}^2}} \qquad V_{cd} \approx \frac{\bar{a} - \bar{b} e^{-i\beta}}{\sqrt{1+\bar{a}^2+C^2}\sqrt{1+\bar{b}^2}} \tag{6a}$$

$$V_{us} \approx \frac{\bar{b} e^{i\beta} - \bar{a}}{\sqrt{1+\bar{a}^2}\sqrt{1+\bar{b}^2+F^2}} \qquad V_{cs} \approx \frac{1}{\sqrt{1+\bar{a}^2+C^2}\sqrt{1+\bar{b}^2+F^2}} \tag{6b}$$

$$V_{ub} \approx \frac{\bar{a}(C-F)}{\sqrt{1+\bar{a}^2}\sqrt{1+F^2}} \qquad V_{cb} \approx \frac{-(C-F)}{\sqrt{1+\bar{a}^2+C^2}\sqrt{1+F^2}} \tag{6c}$$

$$V_{td} \approx \frac{-\bar{b} e^{-i\beta}(C-F)}{\sqrt{1+C^2}\sqrt{1+\bar{b}^2}} \qquad V_{ts} \approx \frac{(C-F)}{\sqrt{1+C^2}\sqrt{1+\bar{b}^2+F^2}} \tag{6d}$$

and

$$V_{tb} = 1 - \mathcal{O}(\tilde{b}^4) . \tag{6e}$$

## $\mathcal{CP}$-Violation in the Mixing Matrix

$\mathcal{CP}$-violation in the Standard Model involves four quark vertices, and all quark processes exhibiting possible violations depend on the representation-invariant quantity $J(a'^2, b'^2, c'^2)$, where[1]

$$-4J^2(a'^2, b'^2, c'^2) = \lambda(a'^2, b'^2, c'^2) = a'^4 + b'^4 + c'^4 - 2a'^2 b'^2 - 2b'^2 c'^2 - 2c'^2 a'^2 . \tag{7}$$

The quantity $\lambda(a'^2, b'^2, c'^2)$ is $-16\mathcal{A}^2$, where $\mathcal{A}$ is the area of the triangle with sides of length $a'$, $b'$ and $c'$. The row formulation has[1]

$$a' = |V_{\alpha j} V_{\alpha k}|, \qquad b' = |V_{\beta j} V_{\beta k}| \quad \text{and} \quad c' = |V_{\gamma j} V_{\gamma k}| \tag{8}$$

where $\alpha$, $\beta$ and $\gamma$ are any permutation of $u$, $c$ and $t$, and where $j$ and $k$ are any two of $d$, $s$ and $b$. Stationary points for $J$ obtained from $(\partial J/\partial \nu^2) = 0$, $\nu = a', b', c'$ lead to right-angled triangles with

$$a'^2 = b'^2 + c'^2, \qquad b'^2 = c'^2 + a'^2 \quad \text{or} \quad c'^2 = a'^2 + b'^2 \ . \tag{9}$$

Given any two sides of the triangle, these conditions correspond to having maximal area $\mathcal{A}$ and hence maximal $J^2$. Taking $\alpha, \beta, \gamma = u, c, t$ respectively and $j = d$, $k = b$ in Equations (8)

$$a' = |V_{ud}V_{ub}| \approx \frac{\bar{a}(C-F)}{(\sqrt{1+\bar{a}^2})^2 \sqrt{1+\bar{b}^2}\sqrt{1+F^2}}$$

$$b' = |V_{cd}V_{cb}| \approx \frac{|\bar{a}-\bar{b}e^{-i\beta}|(C-F)}{(\sqrt{1+\bar{a}^2+C^2})^2 \sqrt{1+\bar{b}^2}\sqrt{1+F^2}} \tag{10}$$

$$c' = |V_{td}V_{tb}| \approx \frac{\bar{b}(C-F)}{\sqrt{1+C^2}\sqrt{1+\bar{b}^2}} \ .$$

We have then that $b'^2 \approx c'^2 + a'^2$ if $\beta = \pm\frac{\pi}{2}$ – i.e., if $D$ is pure imaginary in Equation (1). This extremal condition for $J$ is consistent with experimental results for the Cabibbo angle[2] when one compares these with our expressions (6) for $V_{\alpha i}$ with $\alpha = u, c$ and $i = d, s$ used in conjunction with Equations (3).

**Maximal $\mathcal{CP}$-Violation and the Wolfenstein Parameters**

We consider now the Wolfenstein approximation[3] to the quark mixing matrix

$$V_W = \begin{pmatrix} 1 - \frac{1}{2}\lambda^2 & \lambda & \lambda^3 A(\rho - i\eta) \\ -\lambda & 1 - \frac{1}{2}\lambda^2 & \lambda^2 A \\ \lambda^3 A(1 - \rho - i\eta) & -\lambda^2 A & 1 - \mathcal{O}(\lambda^4) \end{pmatrix} \ . \tag{11}$$

We have then

$$a' = |V_{ud}V_{ub}| \approx \lambda^3 A \sqrt{\rho^2 + \eta^2}$$

$$b' = |V_{cd}V_{cb}| \approx \lambda^3 A \tag{12}$$

$$c' = |V_{td}V_{tb}| \approx \lambda^3 A \sqrt{(1-\rho)^2 + \eta^2}$$

and the conditions (9) lead to

$$
\begin{aligned}
&(i) \quad a'^2 = b'^2 + c'^2 \quad \text{yielding} \quad && \rho = 1 \ . \\
&(ii) \quad b'^2 = c'^2 + a'^2 \quad \text{yielding} \quad && \eta^2 = \rho(1-\rho) \ . \\
&(iii) \quad c'^2 = a'^2 + b'^2 \quad \text{yielding} \quad && \rho = 0 \ .
\end{aligned}
\qquad (13)
$$

The condition $(i)$ implies that the usual unitary triangle of Figure 1 reduces to a straight line with $\eta = 0$.

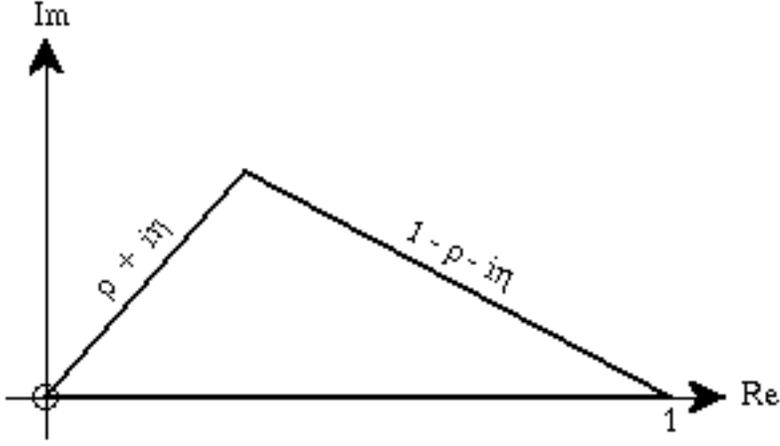

Figure 1: the unitary triangle.

Since in this approximation

$$J = (\lambda^3 A)^2 \eta \qquad (14)$$

this leads to no appreciable $\mathcal{CP}$-violation. Conditions $(ii)$ and $(iii)$ both lead to right-angled triangles in Figure 1. The model described by Equations (1) and (2) yields

$$\left|\frac{V_{ub}}{V_{td}}\right|^2 \approx \left(\frac{\tilde{a}}{\tilde{b}}\right)^2 = \frac{\rho^2 + \eta^2}{(1-\rho)^2 + \eta^2} \ . \qquad (15)$$

Defining $r \equiv (\tilde{a}/\tilde{b})$ leads then to

$$\eta^2 = \frac{r^2}{1-r^2}(1-2\rho) - \rho^2 \qquad (16)$$

which together with result $(ii)$ in Equation (13)

$$\eta^2 = \rho(1-\rho)$$

yields

$$\eta = \pm\frac{r}{1+r^2} \quad \text{and} \quad \rho = \frac{r^2}{1+r^2} \qquad (17a)$$

so that
$$\rho = \pm r\eta \ . \tag{17b}$$

One can also get the results (17a), (17b) by transforming the Kobayashi-Maskawa (KM) mixing matrix given by Equations (6) with $\beta = -\frac{\pi}{2}$ using $V' = SVS^{-1}$ with

$$S = \begin{pmatrix} -e^{-i\psi} & 0 & 0 \\ 0 & 1 & 0 \\ 0 & 0 & 1 \end{pmatrix} \tag{18a}$$

where
$$\cos\psi = \frac{\tilde{a}}{\sqrt{\tilde{a}^2+\tilde{b}^2}} \quad \text{and} \quad \sin\psi = \frac{\tilde{b}}{\sqrt{\tilde{a}^2+\tilde{b}^2}} \tag{18b}$$

in this way getting the positive sign for $\eta$ in (17a). The accuracy of our approximations and of the Wolfenstein matrix in Equation (11) is not justified to order $r^3$ for $\eta$, and therefore

$$\eta \approx r = \sqrt{\frac{m_u}{m_c} \Big/ \frac{m_d}{m_s}} \tag{19a}$$

and
$$\rho \approx r^2 \ . \tag{19b}$$

It follows that since[2] $\frac{m_u}{m_c} \sim \frac{1}{10} \frac{m_d}{m_s}$, the value of $\eta \sim \frac{1}{3}$.

**Maximal $\mathcal{CP}$-Violation in a Special Class of Models**

We consider now a sub-class of mass matrices of the form of Equation (1), with

$$M_U = \Lambda_U \begin{pmatrix} 0 & g_U a^3 e^{i\alpha} & 0 \\ g_U a^3 e^{-i\alpha} & a^2 & C \\ 0 & C & 1 \end{pmatrix} \ , \ M_D = \Lambda_D \begin{pmatrix} 0 & g_D b^3 e^{i\beta} & 0 \\ g_D b^3 e^{-i\beta} & b^2 & F \\ 0 & F & 1 \end{pmatrix} \tag{20}$$

where $g_U$, $g_D$ are of order unity.

These matrices satisfy the inequality conditions of Equations (2) if $a$, $b$, $C$ and $F$, which are taken as real, are considerably less than unity in magnitude, with $C^2 \ll a^2$ and $F^2 \ll b^2$. We take $a$, $b$, $g_U$ and $g_D$ as positive. Maximal $\mathcal{CP}$-violation within this framework, as discussed above, requires $\beta - \alpha = \pm\frac{\pi}{2}$. Models of this form have been discussed in various contexts in References (4), (5) and (6). They yield an approximate geometric progression for the quark masses.

It follows from Equations (7) and (10) with the angle $\beta = \pm\frac{\pi}{2}$

$$|J| \approx \tilde{a}\tilde{b}(F-C)^2 \tag{21a}$$

where
$$\tilde{a} = (a^3 g_{\mathrm{U}})/(a^2 - C^2) \quad \text{and} \quad \tilde{b} = (b^3 g_{\mathrm{D}})/(b^2 - F^2) \tag{21b}$$
using Equations (3) so that
$$|J| = FC(F-C)^2 \left[\frac{y^3 g_U}{y^2 - 1}\right]\left[\frac{z^3 g_D}{z^2 - 1}\right] \tag{22a}$$
where
$$y = \frac{a}{C} \quad \text{and} \quad z = \frac{b}{F} . \tag{22b}$$

We assume – and we discuss this further below – that $g_U = g_U(y)$, $g_D = g_D(z)$ and that $g_U = g_D = g$. We have then that
$$|J| = F^4 x(1-x)^2 \left[\frac{y^3 g(y)}{y^2 - 1}\right]\left[\frac{z^3 g(z)}{z^2 - 1}\right] \tag{23}$$
where $x = \frac{C}{F}$. The maximum in $|J|$, if one exists, as a function of $y$ and $z$ will yield $y = z$. In Reference (5), the result $y = z$ follows from symmetry conditions in the gauge structure of the theory presented there. The nature of mass spectra for down versus up quarks requires $a^2 \ll b^2$ and therefore $C < F$ if $z = y$. The maximum in $|J|$ as a function of $x$, for $x < 1$, comes at $x = \frac{1}{3}$. In the above we have fixed $F$, the largest parameter in our mass matrices (1).

We have from the above, then, that for $y = z$
$$\frac{a}{b} = \frac{C}{F} = x = \frac{1}{3} . \tag{24}$$

To keep the approximate geometric progression property of the quark masses, we should have $g(y) \sim 1$ at the physical value of $y$. Taking $g(y) = 1$, we get the following mass matrices from Equations (20) and (22b) for $y = z$
$$M_U = \Lambda_U \begin{pmatrix} 0 & a^3 & 0 \\ a^3 & a^2 & a/y \\ 0 & a/y & 1 \end{pmatrix} \quad , \quad M_D = \Lambda_D \begin{pmatrix} 0 & -ib^3 & 0 \\ ib^3 & b^2 & b/y \\ 0 & b/y & 1 \end{pmatrix} . \tag{25}$$

Taking $b = 3a = 0.21$ and $y = 4$ yields the following fit to the KM matrix absolute values
$$|V_{KM}| = \begin{pmatrix} 0.975291 & 0.220908 & 0.00280085 \\ 0.220805 & 0.974615 & 0.0370178 \\ 0.0072755 & 0.0364037 & 0.999311 \end{pmatrix} . \tag{26}$$

Using the above values for $b$ and $y$ (or $F$) and suitable values for $\Lambda_U \sim m_t$ and $\Lambda_D \sim m_b$ yields excellent quark mass spectra, approximately given by Equations 2. The value for the $\mathcal{CP}$-violating parameter

$$J \approx \left[\frac{y}{y^2-1}\right]^2 x(1-x)^2 b^4 \approx 2 \times 10^{-5} \ . \tag{27}$$

**Summary**

We have presented here a framework and a class of models with texture zeroes constrained by:

(1) an approximate geometric progression of masses for the quarks,

(2) a certain symmetry in structure for the up and down mass matrices, and

(3) maximal $\mathcal{CP}$-violation consistent with constraints (1) and (2).

Consequences of making $|J|$ extremal are that the $\mathcal{CP}$-violating phase $\beta = \frac{\pi}{2}$, that the spacing of masses for up quarks relative to down quarks is $\sim (b/a)^2 = 9$ times as great, and that the $\mathcal{CP}$-violating parameter $J \sim 2 \times 10^{-5}$. This is, of course, much smaller than the maximum of $J$ without constraints[7,8], which is $(6\sqrt{3})^{-1}$.

The fit to the KM matrix above is considered to be at a scale of $\sim 1$ GeV. These results can be scaled up to the mass of the top quark (for instance) using renormalization group considerations for the Standard Model. From there on the evolution would depend upon the particular theory. Of course, if one had a particular theory in mind, the symmetry breaking would be at some scale above that of the Standard Model and one would then want to evolve down.